\documentclass[a4paper,12pt]{article}

\usepackage{cite}

\usepackage{fullpage}
\usepackage[font=footnotesize,labelfont=bf,margin=1cm]{caption}

\usepackage{color}
\usepackage{amsfonts}
\usepackage{amsmath}
\usepackage{amssymb}

\usepackage{multirow}
\usepackage{multicol}

\numberwithin{equation}{section}
\def    \bea    {\begin{eqnarray}}
\def    \eea    {\end{eqnarray}}

\usepackage{color}
\usepackage{amsfonts}
\usepackage{amsmath}
\usepackage{amssymb}

\usepackage{multirow}
\usepackage{multicol}

\numberwithin{equation}{section}
\newcommand{\beq}{\begin{equation}}
\newcommand{\eeq}{\end{equation}}

\begin{document}

\begin{flushright}
QMUL-PH-22-03
\end{flushright}

\bigskip

\bigskip

\bigskip

\begin{center}
   \baselineskip=16pt
   	{\Large \bf Double copying Exceptional Field theories}
   	\vskip 2cm
   	{\sc		David S. Berman\footnote{\tt d.s.berman@qmul.ac.uk}, Kwangeon Kim\footnote{\tt kim64656@yonsei.ac.kr}, Kanghoon Lee\footnote{\tt kanghoonlee1@gmail.com} }
	\vskip .6cm
    {\small \it 1. Queen Mary University of London, Centre for Theoretical Physics, \\
             School of Physical and Chemical Sciences, London, E1 4NS, England} \\ 
 	\smallskip
    {\small \it 2. Department of Physics, Yonsei University, Seoul 03722, Korea} \\ 
	\smallskip
 {\small \it 3. Asia Pacific Center for Theoretical Physics, Postech, Pohang 37673, Korea} \\
	\vskip 2cm
\end{center}

\begin{abstract}
 We examine exceptional field theory through the lens of the generalised double copy formalism. This allows us to construct classical solutions in M-theory using a generalised Kerr-Schild ansatz and along the way indicates hints towards a single copy of M-theory. Based on a talk at the Nankai Symposium given by DSB.

\end{abstract}

\setcounter{footnote}{0}

\bigskip
\bigskip

\section{Introduction}

The goal of Exceptional field theory (ExFT) is to provide a reformulation of eleven dimensional supergravity while making manifest the exceptional symmetry acting on supergravity fields. This reformulation uses generalised geometry to combine the metric and p-form potentials into a single geometric object. 
For a recent review of the topic and a complete set of references see \cite{Berman:2020tqn}. This reformulation has led to various insights and reinterpretations of M-theory. In particular, M-theory branes such as the membrane and fivebrane are to be viewed as superpositions of waves and gravitational monopoles in the exceptional space as described in \cite{Berman:2014hna, Berman:2014jsa, Berkeley:2014nza}. One successful approach to studying ExFT is to generalise methods from general relativity and adapt them to ExFT. A very useful technique in general relativity that allows us to find solutions of Einstein's equations is the Kerr-Schild ansatz. The Kerr-Schild ansatz has the extraordinary property that it linearises the Einstein equations and so solutions of the resulting linear equations will give an exact solutions to the non-linear Einstein equation! In fact, the resulting linear equations that come from the KS ansatz are Maxwell's equations. This implies the Kerr-Schild ansatz provides us with a map between solutions of Maxwell's equations and solutions in general relativity. This map has been used to formulate a classical solution form of the so called ``double copy" that maps between gravity and Yang-Mills theory \cite{Monteiro:2014cda,Luna:2015paa,Bahjat-Abbas:2017htu,Berman:2018hwd,Huang:2019cja,Alawadhi:2019urr,Kim:2019jwm}. Heuristically, the double copy construction formulates gravity as coming from two copies of Yang-Mills theory as originated in a study of the scattering amplitudes of the two theories \cite{Bern:2010ue,Bern:2010yg}.

We can now ask, instead of just gravity, what if we have a supergravity with additional p-form potentials. What is the single copy of that theory? Very concretely we can ask how we can generalise the Kerr-Schild ansatz in general relativity to include these p-form fields. Answering this is quite hard given the nonlinear nature of Einstein's equations. The key to solving this is to use the reformulation of supergravity in terms of ExFT and then one can use a simple generalisation of the Kerr-Schild ansatz for the generalised metric. This produces again a set of linear equations but this time it is a combination of a Maxwell field and a two form potential. This hints that the single copy of M-theory would be a some product of these one form and two form theories.

\section{$SL(5)$ Exceptional field theory}
We willl work with the exceptional geometry associated 4 dimensions whose coordinates we label $x^\mu$ where $\mu=0,1,2,3$. We will make manifest the U-duality group $SL(5)$. To do this we will extend the dimensionality of the spacetime by including new coordinates that describe membrane wrapping modes. There will be six such coordinates which we call $y_{\mu \nu}$ giving ten in total. A generalised tangent vector $V^{M}$ with $M=1..10$ in this extended space lies in the  $\mathbf{10}$ representation of $SL(5)$ and is parametrised by both the 4-vector $v^\mu$ and the two-form ${\lambda_{\mu \nu}}$ in 4d as follows:
\begin{equation}
  V^{M} = \begin{pmatrix} v^{\mu} \\ \lambda_{\mu\nu}\end{pmatrix} \,,
\label{}\end{equation}
For a more in depth review see \cite{Berman:2010is,Berman:2019biz,Berman:2020tqn}.
We may also represent the $\mathbf{10}$ of $SL(5)$ ($M,N=1\cdots10$) as antisymmetric pairs of $\mathbf{5}$ indices, $[mn]$, where $m, n = 0,1,2,3 $ and $5$, so that
\begin{equation}
\begin{aligned}
   V^{[mn]} &= \begin{cases} V^{\mu 5} = -V^{5\mu} = v^{\mu} \\ V^{\mu\nu} = \tilde{\lambda}^{\mu\nu} = \frac{1}{2} \epsilon^{\mu\nu\rho\sigma} \lambda_{\rho\sigma}\end{cases}
\end{aligned}\label{}
\end{equation}

A key goal of the construction is to combing the metric and 3-form potential into single geometric object called the generalised metric. For the $SL(5)$ theory, the generalized metric $\mathcal{M}_{MN}$ is gvien as follows:
\begin{equation}
\begin{aligned}
  \mathcal{M}_{M N} &= |g|^{\frac{1}{5}}\begin{pmatrix}  g_{\mu \nu} +\frac{1}{4} C_{\mu \rho_{1} \rho_{2}} g^{\rho_{1} \rho_{2}, \sigma_{1} \sigma_{2}} C_{\sigma_{1} \sigma_{2} \nu} & ~~\frac{1}{2} C_{\mu \rho_{1} \rho_{2}} g^{\rho_{1} \rho_{2}, \nu_{1} \nu_{2}} \\ \frac{1}{2} g^{\mu_{1} \mu_{2}, \rho_{1} \rho_{2}} C_{\rho_{1} \rho_{2} \nu} &~~ g^{\mu_{1} \mu_{2}, \nu_{1} \nu_{2}}\end{pmatrix}\,.
\end{aligned}\label{para_genM10-1}
\end{equation}

Along with this generalised metric there will be an ordinary metric $G_{ij}$ with $i,j=1..7$ corresponding to the remaining seven Riemannian dimensions of eleven dimensional supergravity. There will also be additional mixed objects, metric components with legs in the four dimensional space and seven dimensions and components of the three form with legs in the seven dimensional space. We will ignore these in what follows. The action that will describe the dynamics for the fields in question is given by:

\begin{equation}
\begin{aligned}
  S &=\int_{\Sigma} \mathrm{d}^{7}z\, \mathrm{d}^{10}X \sqrt{|G|}~\Big[\ R[G_{ij}] + \frac{1}{12} G^{ij} \partial_{i} \mathcal{M}_{M N} \partial_{j} \mathcal{M}^{M N} -V[\mathcal{M},G]\ \Big]
\end{aligned}
\end{equation}
where $R[G_{ij}]$ is the Ricci scalar with respect to $G_{ij}$, and the scalar potential $V[\mathcal{M},G]$ and $V[m,G]$ are given by

\begin{equation}
\begin{aligned}
 V[\mathcal{M},G] &= -\frac{1}{12} \mathcal{M}^{MN} \partial_{M} \mathcal{M}^{KL} \partial_{N}\mathcal{M}_{KL} + \frac{1}{2} \mathcal{M}^{MN} \partial_{M}\mathcal{M}^{KL} \partial_{K}\mathcal{M}_{LN}  \\
  & -\frac{1}{2} \partial_{M} \mathcal{M}^{MN}\partial_{N} \ln |G| \\
&-\frac{1}{4} \mathcal{M}^{M N}\big(\partial_{M} G_{ij} \partial_{N} G^{ij} + (\partial_{M} \ln |G|) (\partial_{N} \ln |G|)\ \big) \,.
\end{aligned}\label{}
\end{equation}

The variation of the action with respect to the generalized metric $\mathcal{M}$ gives
\begin{equation}
  \delta_{\mathcal{M}} S = \int_{\Sigma} \mathrm{d}^{7}z\, \mathrm{d}^{10}X \sqrt{|G|}~ \delta\mathcal{M}^{MN} \mathcal{K}_{MN}\,,
\label{}\end{equation}
where
\begin{equation}
\begin{aligned}
  \mathcal{K}_{MN}&=- \frac{1}{6\sqrt{|G|}}\partial_{P} \big(\sqrt{|G|}\mathcal{M}^{PQ} \partial_{Q} \mathcal{M}_{MN}\big)+\frac{1}{\sqrt{|G|}} \partial_{P} \big(\sqrt{|G|}\mathcal{M}^{PQ} \partial_{(M} \mathcal{M}_{N)Q}\big) 
  \\
  &\quad +\frac{1}{12} \partial_{M} \mathcal{M}_{PQ} \partial_{N} \mathcal{M}^{PQ} 
   +\frac{1}{6} \mathcal{M}^{PQ} M^{RS} \partial_{P} \mathcal{M}_{R M} \partial_{Q} \mathcal{M}_{S N} 
  \\
  &\quad - \frac{1}{2} \mathcal{M}^{PQ} \mathcal{M}^{RS} \partial_{P} \mathcal{M}_{R M} \partial_{S} \mathcal{M}_{QN} - \frac{1}{2} \partial_{M}\partial_{N}\ln{|G|} +\frac{1}{4} \partial_{M} G_{ij} \partial_{N}G^{ij}
  \\
  &\quad - \frac{1}{6\sqrt{|G|}}\partial_{i} \big(\sqrt{|G|}G^{ij} \partial_{j} \mathcal{M}_{MN}\big)\,.
\end{aligned}\label{}
\end{equation}

To get a consistent variation of $\mathcal{M}$ that is compatible with the coset structure of the generalised metric we need a projection operator $P_{MN}{}^{PQ}$. The equations of motion are then
\begin{equation}
  \hat{\mathcal{R}}_{MN} = P_{MN}{}^{PQ} \mathcal{K}_{PQ} = 0\,,
\label{eom1}\end{equation}

with
\begin{equation}
  P_{MN}{}^{KL} = \frac{1}{\alpha}\left(\delta_{M}^{(K} \delta_{N}^{L)}-\omega \mathcal{M}_{M N} \mathcal{M}^{K L}-\mathcal{M}_{M Q} Y^{Q(K}{}_{ R N} \mathcal{M}^{L) R}\right) \, .
\label{}\end{equation}

The constants $\alpha$ and $\omega$ depend on the theory and are as in the following table:
\begin{table}[h!]
\centering
\begin{tabular}{cccc} $G$ & $H$  & $\alpha$ & $\omega$ \\ \hline $GL(d)$ & $SO(d)$ & $1$ & 0 \\ $O(D, D)$ & $O(D) \times O(D)$ & $2$ & 0 \\ $SL(5)$ & $SO(5)$ &  3 & $-\frac{1}{5}$ \\ $Spin(5,5)$ & $Spin(5) \times Spin(5)$ &  4 & $-\frac{1}{4}$ \\ $E_{6(6)}$ & $USp(8)$ &  6 & $-\frac{1}{3}$ \\ $E_{7(7)}$ & $SU(8)$ &  $12$ & $-\frac{1}{2}$ \\ $E_{8(8)}$ & $SO(16)$ &  $60$ & $-1$
\end{tabular}
\caption{List of duality groups $G$ and their maximal compact subgroups $H$ (Euclidean case) the generalised metric parameterises $G/H$}
\label{table:1}\end{table}

\newpage

\section{Generalised Kerr-Schild}
We now present the generalised Kerr-Schild ansatz for the generalised metric that we will use to solve the equations of motion \cite{Berman:2020xvs}:

\begin{equation}
\begin{aligned}
  \mathcal{M}_{MN} &= \mathcal{M}_{0 MN} +\kappa \varphi P_{0MN}{}^{PQ} K_{P} K_{Q}\,,
  \\
  \big(\mathcal{M}^{-1}\big){}^{MN} &= \big(\mathcal{M}^{-1}_{0}\big){}^{MN} - \kappa\varphi P_{0}{}^{MN}{}_{PQ} K^{P} K^{Q}\,.
\end{aligned}\label{Gen_KS_ansatz}
\end{equation}

$ \mathcal{M}_{0 MN}$ is the background generalised metric which we take to be flat. $K_{M}$ is a null vector with respect to $\mathcal{M}_{0}$, which means:
\begin{equation}
K_{M} \big(\mathcal{M}_{0}^{-1}\big){}^{MN} K_{N} = 0 \, .
\end{equation}
Finally, $\varphi$ is a scalar field that must be determined. 

All the indices are raised and lowered by the background generalised metric $\mathcal{M}_{0}$ and $P_{0MN}{}^{PQ}$ is constructed with the  trivial background generalised metric. The null condition on $K_{M}$ is not sufficient to linearise the inverse metric. 
To do this so we need an additional constraint, defining:
\begin{equation}
Q_{MN} = \varphi P_{0MN}{}^{PQ} K_{P} K_{Q} \,
\end{equation}
we need
\begin{equation}
  Q_{MP} Q^{PQ} = 0\, .
\label{nil_Q}\end{equation}
In general relativity the projection operator is trivial, 
\begin{equation}
  \big(P^{\scriptscriptstyle \mathrm{GR}}\big){}_{\mu\nu}{}^{\rho\sigma} = \delta_{\mu}{}^{(\rho} \delta_{\nu}{}^{\sigma)}\,.
\label{}\end{equation}
and so from the form \eqref{Gen_KS_ansatz}, the Kerr-Schild ansatz for relativity is given by 
\begin{equation}
\begin{aligned}
    g_{\mu\nu} &= \tilde{g}_{\mu\nu} + \kappa \varphi K_{\mu} K_{\nu}\,,
    \\
    \big(g^{-1}\big){}^{\mu\nu} &= \big(\tilde{g}^{-1}\big){}^{\mu\nu} - \kappa \varphi K^{\mu} K^{\nu}\,,
\end{aligned}\label{}
\end{equation}
where $\tilde{g}_{\mu \nu}$ is the background metric which the vector field $K^{\mu}$ is null with respect to:
\begin{equation}
K^\mu \tilde{g}_{\mu\nu} K^\nu =0 \, .
\end{equation}
The Einstein equations then reduce to Maxwell's equations for $\varphi K_{\mu}$ and so this allows us to identify  $\varphi K_{\mu}$ as the single copy Maxwell field, $A_\mu$.

Before moving on to the $SL(5)$ exceptional case, let us do the Kerr Schild ansatz for Double Field Theory as first reported in \cite{Lee:2018gxc}. For Double Field Theory, the generalised metric $\mathcal{H}_{MN}$ on a $2D$ dimensional doubled space is given by the coset $\mathit{O}(D,D)/\mathit{O}(1,D-1)\times \mathit{O}(1,D-1)$. In terms of the usual $D$ dimensional metric $g$ and the Kalb-Ramond two-form $B$ it is written as follows:
\begin{equation}
  \mathcal{H}_{MN}=\begin{pmatrix}  g_{\mu\nu} -  B_{\mu\rho} g^{\rho\sigma} B_{\sigma\nu} & B_{\mu\rho}g^{\rho\nu} \\ -g^{\mu\rho}B_{\rho\nu} & g^{\mu\nu} \end{pmatrix}\, . \label{eq:genmetric}
\end{equation}
It is a symmetric $\mathit{O}(D,D)$ element and satisfies the $\mathit{O}(D,D)$ compatibility constraint, $\mathcal{H}_{MP} \mathcal{J}^{PQ} \mathcal{H}_{QN} = \mathcal{J}_{MN}$, where $\mathcal{J}_{MN}$ is the $\mathit{O}(D,D)$ metric which maybe chosen to be:
\begin{equation}
  \mathcal{J}_{MN} = \begin{pmatrix} \mathbf{0} & \delta^{a}{}_{b} \\ \delta_{a}{}^{b} & \mathbf{0} \end{pmatrix}\,.
\label{}\end{equation}

The projector uses the so called $Y$-tensor \cite{Berman:2012vc} which is given in terms of the $\mathit{O}(D,D)$ metric by
\begin{equation}
  Y^{PQ}{}_{MN}= \mathcal{J}^{PQ} \mathcal{J}_{MN}\, .
\label{}\end{equation}
The projection operator then factorises as follows:
\begin{equation}
  \big(P^{\scriptscriptstyle \mathrm{DFT}}\big){}_{MN}{}^{PQ} = 2 P_{(M}{}^{P} \bar{P}_{N)}{}^{Q}\,,
\label{}\end{equation}
where $P_{M}{}^{N}$ and $\bar{P}_{M}{}^{N}$ are projectors defined by the generalized metric of DFT, $\mathcal{H}_{MN}$, and the $O(D,D)$ metric  $\mathcal{J}_{MN}$:
\begin{equation}
  P_{MN} = \frac{1}{2}\big(\mathcal{J}_{MN} + \mathcal{H}_{MN}\big) \,, \qquad \bar{P}_{MN} =\frac{1}{2}\big(\mathcal{J}_{MN} - \mathcal{H}_{MN}\big)\,.
\label{}\end{equation}
The factorisation of the projector, should not in fact be a surprise, it is in fact a consequence of the left and right decomposition of the closed string modes.

Applying to the general KS ansatz \eqref{Gen_KS_ansatz}, we have the KS ansatz for $\mathcal{H}_{MN}$
\begin{equation}
\begin{aligned}
    \mathcal{H}_{MN} &= \mathcal{H}_{0MN} + \kappa \varphi \big(P^{\scriptscriptstyle \mathrm{DFT}}_{0}\big)_{MN}{}^{PQ} K_{P} K_{Q} \,,
\end{aligned}\label{DFT_KS}
\end{equation}
where $K_{M}$ is a null vector, $K_{M} \mathcal{J}^{MN} K_{N} = 0$ and $\big(P^{\scriptscriptstyle \mathrm{DFT}}_{0}\big)_{MN}{}^{PQ}$ is a background projection operator. It is useful to write the projected null vectors as
\begin{equation}
  L_{M} = P_{0M}{}^{N} K_{N}\,, \qquad \bar{L}_{M} = \bar{P}_{0M}{}^{N} K_{N}
\label{}\end{equation}
and from the completeness relation, $\delta_{M}{}^{N} = P_{M}{}^{N} + \bar{P}_{M}{}^{N}$, $K_{M}$ is decomposed to $L_{M}$ and $\bar{L}_{M}$

$Q_{MN}$ then may written in terms of $L_M, \bar{L}_N$ as follows:
\begin{equation}
  Q_{MN} = \varphi \big(L_{M}\bar{L}_{N} + L_{N}\bar{L}_{M}\big)\,.
\label{}\end{equation}
We require the nilpotency condition on $Q_{MN}$ \eqref{nil_Q} for the linearity of the KS ansatz. This  implies that $L_{M}$ and $\bar{L}_{M}$ have to be mutually orthogonal null vectors of definite chirality and so,
\begin{equation}
  L_{M} = P_{0M}{}^{N} L_{N} \,, \qquad \bar{L}_{M} = \bar{P}_{0M}{}^{N} \bar{L}_{N}
\label{chirality_null}\end{equation}
and
\begin{equation}
  L_{N} L^{N}=0\,, \qquad \bar{L}_{N}\bar{L}^{N} = 0\,.
\label{}\end{equation}
Then the Kerr Schild ansatz is then rewritten as:
\begin{equation}
    \mathcal{H}_{MN} = \mathcal{H}_{0MN} +\kappa \varphi \big(L_{M} \bar{L}_{N} + L_{N} \bar{L}_{M}\big)\,.
\label{}\end{equation}

From this the equations of motion become that of two Maxwell fields from which the B-field and metric solutions may be reconstructed \cite{Lee:2018gxc}.
\bigskip

But what about exceptional field theory?

We will now assume the background generalised metric $\mathcal{M}_{0}$ is:
\begin{equation}
  \mathcal{M}_{0MN} = \begin{pmatrix} \eta_{\mu\nu} &0 \\ 0 & \eta^{\mu\mu',\nu\nu'} \end{pmatrix} \,, \quad  \mbox{and} \quad \mathcal{M}_{0mm',nn'} = \begin{pmatrix} \eta_{\mu\nu} &0 \\ 0 & - \eta_{\mu\mu',\nu\nu'} \end{pmatrix} \,,
\label{flat_gen_metric}\end{equation}
and the associated null vector $K_{M}$ is parametrized as
\begin{equation}
  K_{M} = \begin{pmatrix} l_{\mu} \\ k^{\nu\nu'} \end{pmatrix}\,, \qquad K^{M} = \begin{pmatrix} l^{\mu} \\ k_{\nu\nu'} \end{pmatrix}\,,
\label{}\end{equation}
and
\begin{equation}
  K_{mm'} = \begin{cases}
  K_{\mu5} = l_{\mu}
  \\
  K_{\nu\nu'} = \tilde{k}_{\nu\nu'}
  \end{cases}
  \qquad 
  K^{mm'} = \begin{cases}
  K^{\mu5} = l^{\mu}
  \\
  K^{\nu\nu'} = - \tilde{k}^{\nu\nu'}
  \end{cases}\,.
\label{comp_null}\end{equation}
Here the Greek indices are raised and lowered by the flat background metric $\eta_{\mu\nu}$.

$\tilde{k}_{\mu\nu}$ and $\tilde{k}^{\mu\nu}$ are related with $k_{\mu\mu'}$ and $k^{\mu\mu'}$ as
\begin{equation}
  \tilde{k}_{\mu\mu'} = \frac{1}{2} \epsilon_{\mu\mu'\nu\nu'} k^{\nu\nu'}\,, \qquad \tilde{k}^{\mu\mu'} = \frac{1}{2}\eta^{\mu\mu',\nu\nu'} \tilde{k}_{\nu\nu'}  = -\frac{1}{2} \epsilon^{\mu\mu'\nu\nu'} k_{\nu\nu'}\,,
\label{}\end{equation}
and their inner product satisfies
\begin{equation}
  k_{\mu\mu'} k^{\mu\mu'} = - \tilde{k}_{\mu\mu'} \tilde{k}^{\mu\mu'}\,.
\label{}\end{equation}
Then the null condition for $K$ is then
\begin{equation}
  K_{M} K^{M} = K_{mm'} K^{mm'}= l_{\mu} l^{\mu} + \frac{1}{2} k_{\nu\nu'}k^{\nu\nu'} = l_{\mu} l^{\mu} - \frac{1}{2} \tilde{k}_{\nu\nu'}\tilde{k}^{\nu\nu'} = 0\,.
\label{}\end{equation}
The projection operator $P_{M N}{}^{P Q}$ for  the $SL(5)$ ExFT is:
\begin{equation}
  P_{M N}{}^{{P Q}} = \frac{1}{3}\Big(\delta_{(M}{}^{P} \delta_{N)}{}^{Q} +\frac{1}{5} \mathcal{M}_{MN} \mathcal{M}^{P Q}-\mathcal{M}_{M R} Y^{R(P}{}_{S N} \mathcal{M}_{}{}^{Q) S}\Big)\,,
\label{Projection_SL5}\end{equation}
where the $Y$-tensor is
\begin{equation}
  Y^{MN}{}_{PQ} = Y^{mm'nn'}{}_{pp'qq'} = \epsilon^{mm'nn'r} \epsilon_{pp'qq'r} \, .
\label{}\end{equation}
Combining the above results allows us to write the following Kerr-Schild ansatz for $\mathcal{M}_{MN}$
\begin{equation}
\begin{aligned}
  \mathcal{M}_{MN} &= \mathcal{M}_{0MN} +\kappa\varphi P_{0 M N}{}^{P Q}K_{P}K{}_{Q} \,,
  \\
  (\mathcal{M}^{-1})^{MN}  &= \mathcal{M}_{0}{}^{MN} - \kappa\varphi P_{0}{}^{M N}{}_{P Q}K^{P}K^{Q} \,,
\end{aligned}\label{KS_SL5}
\end{equation}
where $\kappa$ is some constant parameter associated to the appropraite Newton's constant and the indices are raised and lowered by the background generalized metric $\mathcal{M}_{0MN}$ and $(\mathcal{M}^{-1}_{0})^{MN}$ in \eqref{flat_gen_metric}. 

We can now express the ordinary metric and three form potential in terms of the components of $K^M$ as follows:
\begin{equation}
\begin{aligned}
  |g| &= \big|1 -\frac{\kappa \varphi}{3} l\cdot l\big|^{-\frac{5}{3}}  \,,
  \\
  g_{\mu \nu} &= \big|1 -\frac{\kappa \varphi}{3} l\cdot l\big|^{-\frac{2}{3}}  \Big(\eta_{\mu \nu} + \frac{\kappa\varphi}{3} \big(l_{\mu} l_{\nu} - \tilde{k}_{\mu\rho} \tilde{k}_{\nu}{}^{\rho}\big)\Big) \,,
  \\
  C_{\mu \nu \rho} &= \frac{2\kappa\varphi}{3|1 -\frac{\kappa \varphi}{3} l\cdot l|} l_{[\mu} k_{\nu \rho]}\,.
\end{aligned}\label{KS_components2}
\end{equation}
where $l\cdot l = l_{\mu} \eta^{\mu\nu} l_{\nu}$. 
Remarkably, the KS ansatz for the generalized metric is linear in $\kappa$, but component fields are highly nonlinear. If we set $k_{\mu\nu} = 0$, then $l^{\mu}$ becomes a null vector and the KS ansatz reduces to the conventional KS ansatz in GR, $g_{\mu\nu} = \eta_{\mu\nu} + \kappa\varphi l_{\mu} l_{\nu}$ and $C_{\mu\nu\rho} = 0$.

When we insert this ansatz into the equations of motion we find:
\begin{equation}
\begin{aligned}
  &\partial^{\sigma} \partial_{\sigma} \big(\varphi l_{\mu}\big)-\partial^{\sigma} \partial_{\mu} \big( \varphi l_{\sigma}\big) =0\,,
  \\
  &\partial^{\rho} \partial_{[\rho} \big(\varphi k_{\mu\nu]}\big) = 0\,.
\end{aligned}\label{}
\end{equation}
From which we can identify:
\begin{equation}
\begin{aligned}
  A_{\mu} &=  \varphi l_{\mu}\,,   \qquad  B_{\mu\nu} &= \varphi k_{\mu\nu} \,.
\end{aligned}\label{}
\end{equation}
as the single copy fields.

This gives the Maxwell field and two form potential as the single copy for gravity with a three form potential.

As a test of this ansatz lets construct the membrane solution in terms of the generalised Kerr-Schild Ansatz and interpret these solutions from the point of view of the single copy fields. 

Take the worldvolume directions of the M2-brane to be $t, x^{1},x^{2}$ and denote them as $x^{\alpha} = \{t,x^{1},x^{2}\}$ and choose the M-theory circle direction as $x^{2}$. The transverse directions are $\vec{x}_{8} = \{x^{3},z^{i}\}$, where $i,j,\cdots$ are 7-dimensional extra directions. The M2-brane geometry is given by
\begin{equation}
\begin{aligned}
  &\mathrm{d}s^{2}_{11} = H^{-\frac{2}{3}} \eta_{\alpha\beta} \mathrm{d}x^{\alpha} \mathrm{d}x^{\beta} + H^{\frac{1}{3}} \mathrm{d}x^{3} \mathrm{d}x^{3}+ H^{\frac{1}{3}} \delta_{ij} \mathrm{d}z^{i} \mathrm{d}z^{j}\,,
  \\
  &C_{t12} = -\big(1-H^{-1}\big)\,,\qquad H= 1+\frac{h}{|\vec{x}_{8}|^{6}}\,.
\end{aligned}\label{}
\end{equation}

To embed the 11-dimensional supergravity into the $SL(5)$ ExFT, we need to use the following Kaluza-Klein ansatz for the 11-dimensional metric $\hat{g}$ 
\begin{equation}
  \hat{g}_{\hat{\mu} \hat{\nu}}=\left(\begin{array}{cc}|g|^{-\frac{1}{5}} G_{ij} + A_{i}^{\rho} A_{j}^{\sigma} g_{\rho\sigma} & A_{i}^{\rho} g_{\rho\nu} \\  g_{\mu k} A_{j}^{k} & g_{\mu\nu}\end{array}\right)\,,
\label{bigmet}\end{equation}

The internal metric and $C$-field are
\begin{equation}
  g_{\mu\nu} = H^{-\frac{2}{3}} \big( \eta_{\alpha\beta} \mathrm{d}x^{\alpha} \mathrm{d}x^{\beta} + H\mathrm{d}x^{3}\mathrm{d}x^{3}\big)\,, \qquad C_{012} = H^{-1} -1  \, .
\label{}\end{equation}
 The single copy of this solution is then given by: 
\begin{equation}
A_0 = \varphi  \,, \qquad B_{12}= \varphi\,.
\end{equation}
A quite remarkable result. Interpreting in terms of the single copy fields the membrane is described by a two dimensional plane of electric charges. That is the Maxwell field is given by solving the Harmonic equation for an electric source smeared over two spatial dimensions just like the classical electrostatic problem for a plane of charge. The two form solution is like that of a magnetic string that is smeared over one additional dimension to give a plane of string charge.

\section{Conclusions and Discussion}
Whether one appreciates the need for reformulating eleven dimensional supergravity in terms of expectional field theory or not, the generalised Kerr-Schild ansatz allows one to vastly simplify Einstein's equations sourced by a three form potential. It is doubtful that without ExFT one would have the imagination to construct the generalised Kerr-Schild ansatz. This of course can be extended to other fields such as the six form in eleven dimensional supergravity but to do this requires a $E7$ Kerr-Schild ansatz. It is also curious that the single copy produces a Maxwell and a two-form field. Thus indicating that the single copy for M-theory is more complicated than a product of Yang-Mills fields and instead must involve some form of Gerbe.

\section{Acknowledgements}
DSB's work is partial supported by Pierre Andurand.

\addcontentsline{toc}{section}{References}
\bibliographystyle{JHEP} 
\bibliography{CurrentBib}

\end{document}